       \documentstyle  {article}
 \evensidemargin\oddsidemargin
\begin{document}
\count0 = 1
 \title{\small{  QUANTUM MEASUREMENTS AND INFORMATION\\ 
RESTRICTIONS IN ALGEBRAIC QM\\ }}
\small\author{S.N.Mayburov \\ 
Lebedev Inst. of Physics\\
Leninsky Prospect 53, Moscow, Russia, 117924\\
E-mail :\quad   mayburov@sci.lpi.msk.su}
\date {}
\maketitle
\begin{abstract}
It's argued that
  Information-Theoretical restrictions for the systems selfdescription
are important for Quantum Measurement problem.
As follows from Breuer theorem,
for the quantum object S  measurement by information system $O$
they  described by $O$ restricted states $R_O$.  
$R_O$ ansatz  can be introduced phenomenologically from the consistency
with Shr$\ddot {o}$dinger dynamics and measurement statistics.
 The analogous  restrictions obtained in Algebraic QM 
 considering Segal algebra  of S,$O$ observables
and the resulting $O$ algebraic states $\{\varphi^O\}$ set
defined as its dual space. From
 Segal theorem for associative  (sub)algebras  it's shown
 that  $\varphi^O_j$ describes the  random 'pointer'
outcomes $O_j$  observed by $O$ in the individual events.

\end{abstract}
\vspace{6mm}
$ \quad \quad $ {\bf Key Words: \, Quantum Measurements, Information,
  $C^*$ algebras}  
\small { \quad 
\vspace{10mm}
%
%
%

\section  {Introduction}

Despite that Quantum Mechanics (QM) is universally acknowledged
physical theory, there are still several unresolved problems concerned with
its interpretation.
 Of them, the State Collapse or Quantum Measurement
   Problem is the most widely
and long discussed (D'Espagnat,1990; Busch,1996).
In this paper we regard the quantum  measurement process in the
 information-theoretical framework and demonstrate
its importance for the state collapse consideration
(Svozil,1993). Really,   both the quantum and classical measurement
 is, eventually, the  information
acquisition   by the information system $O$ (Observer) 
via the direct or indirect interaction with the studied system S
 (Guilini,1996; Duvenhage,2002).  
Therefore, 
  the possible restrictions on the information
pattern transferred from S to $O$ can be important
in the Measurement Theory (Breuer,1996).
We  concede in our study that QM description is applicable
both for  a microscopic    and  macroscopic
objects; in particular,
 $O$ state described by  Dirac  vector
$|O\rangle$ or density matrix $\rho$ 
  relative to another  observer $O'$ (Rovelli,1995;Bene,2000).
 $O$ considered as the information gaining and utilizing system 
(IGUS)  which acquire and memorize the information as the result of
 S interactions   with the measuring system (MS)  which  element is $O$
(below S formally is also regarded as MS element).
In principle, $O$  can be  either a human brain or some automatic device 
processing the information. In all cases it's
the system with some number of  internal degrees of freedom (DF)  which
 interacts during S information acquisition, so
 that $O$ internal state changes after it.

   S measurement by $O$ described by MS state $|MS\rangle$ evolution
relative to some  $O'$, yet in this model  
 the  acquired S information memorized and processed by $O$, not by $O'$,
 which reflected by $O$ internal state   evolution. Therefore, 
the detailed description of S information recognition    
  should be analyzed in the  selfdescription framework (Svozil,1993).
The information systems selfdescription
was  studied already in the context
of the  selfreference  problem 
 (Finkelstein,1988; Mittelstaedt,1998).
It was shown that the arbitrary system selfdescription
is always incomplete; this result often interpreted as the
analog of G$\it\ddot{o}$del Theorem for Information theory (Svozil,1993).
 In this  framework Breuer developed 
 the  restricted  states formalism
 for the selfdescritpion
 in the measurement process - the  selfmeasurement which is applicable 
 both  in classical and quantum case (Breuer,1996).
 It follows that $O$ internal state $R_O$, which
 is the partial (restricted) MS state,   can differ
 principally from the standard
QM ansatz for $O$ state relative to $O'$ (Mittelstaedt, 1998).
Basing on this  results, we propose here the novel
formalism  which    accounts $O$ selfmeasurement  effects and predicts
 the measured state collapse.
 Its  main feature is the modification of quantum state ansatz which becomes
 the doublet $\Phi= \{ \phi^D, \phi^I \} $, where $\phi^D=\rho$ is  QM
density matrix MS, $\phi^I(n)$ is $O$ restricted state describing $O$
 subjective information 
in the given individual event $n$ 
(Mayburov,2001).
$\phi^I$  can be  independent of $\phi^D$ and ,in particular,
  demonstrates the stochastic behavior in S measurements.  
 It will be shown that such formalism
  corresponds to the well-known generalization
of standard QM - algebraic QM based on Jordan,Segal and $C^*$- algebras
applications (Emch,1972). In its framework $\phi^D$ is MS state defined on MS 
observables algebra $\cal{U}$,
 $\phi^I$ corresponds to the 
  state defined on $O$ observables subalgebra $\cal{U}_O$.


%
We must stress that  the observer consciousness 
 never referred directly and doesn't play any role in our theory 
(London,1939).
 Rather, in our model observer $O$  regarded as the quantum 
reference frame (RF) which interacts with studied object S (Aharonov,1981).
 S state description 'from the point of view' of the particular $O$
referred by the terms 'S state in $O$ RF' or simply 'S  state for $O$'.
 The terms 'perceptions', 'impressions' used by us to characterize the
   IGUS $O$ description of experimental  results 
and   defined below in strictly physical  terms.
In particular, the perception is  
 the acquisition of some information by IGUS, i.e. the change
of IGUS  state; 
the different $O$ impressions  associated with the different, 
 $O$ physical states.

\section {Measurements  and Quantum States Restrictions}

Our formalism exploits both the quantum states in the individual events
- i.e. individual states and the 
  statistical states describing the quantum ensembles properties
 (Mittelstaedt,1998).
 Remind that in QM  the individual
states are the pure states which are isomorphic to Dirac vectors 
 $|\Psi\rangle$    
in $\cal H$ ; the statistical states  described by  the 
normalized, positive operators of trace $1$
- density matrixes  $\rho$ on $\cal {H}$.
 If the  $\Psi_l$ composition  is known  for 
the given ensemble, its state can be described in more detail by
 the   ensemble state (Gemenge)    
 presented  by the table $W^e=\{ \Psi_l; P_l\}$ 
where $P_l$ are the corresponding probabilities (Busch,1996).
   Algebraic QM   states  will be considered in chap. 3.

We'll regard 
  the simple  MS measurement model similar 
to  von Neuman model (von Neuman,1932; Busch,1996).
It     includes
the measured binary state  S which interacts with the
 observer $O$  storing the incoming S information.
 In our model the detector D omitted in  MS chain,
the role of $O$ decoherence effects will be discussed below.
%
The regarded $O$ has one  internal DF  and  in its Hilbert space $\cal {H}_O$
 the basis  consists of the
  three orthogonal states $|O_{0,1,2}\rangle$
 which are  the eigenstates of $Q_O$  
'internal pointer' observable with eigenvalues $q^O_i$.
We'll  consider    
 the measurement   of the binary
 S observable $\hat{Q}$ on S   state $\psi_s$.
Initial  $O$ state  is  $|O_0\rangle$ and  MS initial state is :
\begin {equation} 
     \Psi^{in}_{MS}=\psi_s |O_0\rangle= 
(a_1|s_1\rangle+a_2|s_2\rangle)|O_o\rangle  \label {AAB}
\end {equation}
 where $|s_{1,2}\rangle$ are $Q$ eigenstates with eigenvalues $q_{1,2}$.
  S-$O$ measuring interaction starts at $t_0$
and finished effectively at some finite $t_1$,  
 by the suitable choice of  S$-O$ interaction Hamiltonian $\hat{H}_I$ 
 Schr$\ddot{o}$dinger equation (SE)  results in
 MS final  state $\rho^p_{MS}$ :
\begin {equation}
   \Psi_{MS}=\sum \Psi^{MS}_j = \sum a_i|s_i\rangle|O_i\rangle
                                   \label {AA2}
\end {equation} 
As the result, for any $\psi_s$ one obtains $ \bar{Q}_O=\bar{Q}$
 which means that $O$ performs the unbiased $Q$ measurement.
Meanwhile, for any $O$ observable
 $Q'_O  \ne F(Q_O) ;\, \bar{Q}'_O=0$ independently of 
$\psi_s$. 
Regarding $O$ as the information system, 
 we'll assume  that $|O_{1,2,0}\rangle$ corresponds to 
 $O$ information pattern - an impressions notified by $q^O_{1,2,0}$
(Guilini,1996).
%
%
Therefore,  at $t>t_1$  for external  $O'$ 
MS is  in the pure state  $\Psi_{MS}$ of (\ref {AA2})
 which is  the superposition of the states 
corresponding to the different measurement outcomes.
Basing on our  assumptions, 
 from $O$ 'point of view' $\Psi_{MS}$ describes the
simultaneous superposition (coexistence) of two
contradictory impressions : $Q_O=q^O_1$ and $Q_O=q^O_2$ percepted by $O$
 simultaneously.
 Yet it's well known   that experimentally   
the macroscopic   $O$ observes at random one of $Q_O$ values $q^O_{1,2}$.
 From 
that  S final state is $|s_1\rangle$ or $|s_2\rangle$  and
S state collapse occurs. 
In standard QM  with Reduction Postulate S final  state described by
  the density matrix  of mixed state: 
\begin {equation}
 \rho_s^m= \sum_i |a_i|^2|s_i \rangle \langle s_i|
                                                              \label {AA33}
\end {equation}
In accordance with it, in our model one
 can  ascribe to MS the corresponding mixed state :
\begin {equation}
 \rho_{MS}^m= \sum_i |a_i|^2|s_i \rangle \langle s_i||O_i \rangle \langle O_i|
                                                              \label {AA3}
\end {equation}
which    differs principally from $\rho^p_{MS}$ of (\ref {AA2}). 
It's quite difficult to doubt both in the correctness of
  MS evolution description by SE
 and in  the state collapse experimental observations.
  This  obvious contradiction constitutes famous Wigner 
  'Friend  Paradox' for $O, O'$ (Wigner,1961).
We attempt here to unite  this alternative  
  systems descriptions 'from outside' by $O'$
and 'from inside' by $O$ in the same  formalism.

Formally, both the classical and quantum measurement 
 of the arbitrary  system $S'$ is the 
mapping of $S'$ states set $N_S$ on  the given IGUS $O^S$ states set $N_O$ 
(Mittelstaedt,1998).  
 If the final $O^S$ and $S'$ state can't be factorized,
 then  $O^S$ should be regarded as
 the subsystem of the large system $S_T=S'+O^S$ with the states set $N_T$.
  In this situation - 'measurement from inside'  $N_O$ is $N_T$ subset and 
$O^S$ state  is $S_T$ state projection to $N_O$ -
  the restricted  state $R_O$.
From $N_{T}$ mapping properties the principal restrictions on
 $O^S$ restricted states obtained in
 Breuer theorem : if for two arbitrary $S_T$ 
states $\Phi_{S}, \Phi'_{S}$ 
their restricted  states $R_O, R'_O$ coincide, then for $O^S$ this $S_T$ 
states are indistinguishable (Breuer,1996).
  The origin of this  results  in classical case is easy to 
 understand: $O^S$ has less number
 of DFs then $S_T$ and, therefore,
 can't describe completely $S_T$ state (Svozil,1993).
 In quantum case  
  the  observables noncommutativity and nonlocality introduce
some new features regarded below.
Despite that $R_O$ are incomplete $S_T$ states,
 they are the real physical states
for $O^S$ observer - 'the states in their own right' as Breuer puts it. 

 The described
$S'$,$O^S$,$S_T$ relation corresponds to our MS model which can be
regarded as 'the MS measurement from inside'. 
  Breuer results doesn't permit
to derive the restricted states for an  arbitrary  system
directly,
 and as the phenomenological $R_O$ ansatz it was proposed (Breuer,1996)  
to use the partial trace  which for MS final  state  (\ref {AA2}) is equal to:
\begin {equation} 
   R_O=Tr_s  {\rho}^p_{MS}=\sum |a_i|^2|O_i\rangle\langle O_i|
      \label {AA4}
\end {equation}
%
%
%
in particular, for the incoming  $|s_j\rangle$  
 $R_O=|O_j\rangle \langle O_j|$.
For   MS   state $\rho^{m} _{MS}$
of (\ref {AA3}) appearing in the measurement of the incoming S mixture,
 the  corresponding restricted statistical
 state is the same $R^{mix}_O=R_O$.
This equality doesn't mean  the
 collapse of MS pure state $\Psi_{MS}$
 because  the collapse appearance
   should be   verified also  for  MS,$O$ individual
  states.
For the pure case MS individual state is always $\Psi_{MS}$, yet
 for  the incoming S statistical mixture (\ref {AA3})
  MS individual state   differs
from event to event:
\begin {equation}
\rho^A (n)=\rho^I_l=|O_l\rangle \langle O_l|| s_l\rangle\langle s_l|
  \label {A44}
\end {equation}
where the random $l(n)$ described by the
 probabilistic distribution $P_l=|a_l|^2$.
 $\rho^A (n)$  differs from
 the  state (\ref{AA2}), correspondingly,
its restricted state 
 $\varsigma^O(n)=|O_l\rangle \langle O_l|$
  also differs in any event  from $R_O$ of (\ref {AA4}). Due to it, 
 the main condition of Breuer  Theorem violated for the individual states
 and $O$ can differentiate pure/mixed states 'from inside'
in the individual events (Breuer,1996). Therefore, the proposed 
formalism doesn't permit to obtain 
 the state collapse  for $O$ selfdescription
 in standard QM framework.
Hence, $R_O$ is the consistent restriction of MS statistical state
$ \rho_{MS}^{p,m}$ to $O$  which coincides  for the
pure and mixed  S states with the same $|a_i|$. 
 $R_O$ ansatz (\ref {AA4})
 regarded also as $O$ individual partial state relative to 
external classical observer  in the standard Quantum Measurement
Theory without selfdescription  (Lahti,1990). 


Note that even  in Breuer  theory 
 $O$ can't observe the difference between 
  MS   states with different $D_{12}=a_1^*a_2+a_1 a_2^*$. 
 Such difference revealed by 
  MS interference term (IT) observable :
\begin {equation}
   B=|O_1\rangle \langle O_2||s_1\rangle \langle s_2|+j.c.
    \label {AA5}
\end {equation}
In standard QM,
being measured by external $O'$ on S,$O$,
 it gives $\bar{B}=0$ for the mixed MS state (\ref {AA3}),
 but  $\bar{B}\neq 0$  for the pure MS states (\ref{AA2});
 $B$ value principally can't be measured by $O$ 
'from inside';
 note also that $B,Q_O$  doesn't commute. 
%
%
%

Formally, MS individual state  for $O$ can be written  in doublet 
form $\Phi^B(n)=|\phi^D,\phi^I \gg$, where $\phi^D=\rho_{MS}$
is the objective (dynamical)
state component  and the information component
 $\phi^I$ describes $O$ subjective information  in the given event $n$.
 In Breuer theory  
 for the pure MS  states $\phi^I$ is just $\phi^D$  projection
but in the alternative formalism described below it will
 describe the novel $O$ state features.
%
%
%
In this formalism the state collapse 
appears in MS  'measurement from inside' performed by $O$ and
reflected in its information component $\phi^I$ (Mayburov, 2001).
%
To agree with the
 quantum Schr$\rm\ddot{o}$dinger dynamics  (SD),
 the particular formalism should satisfy to two operational conditions : \\
i) if an arbitrary system $S'$   doesn't interact with IGUS $O^S$,
 then for $O^S$ this  system evolves according to
 Schr$\rm\ddot{o}$dinger-Liouville  equation  (SLE)  \\
ii) If $S'$ interacts with $O^S$ and the entangled $S'$,$O^S$ state produced
 i.e. measurement occurs, then  SD can be violated for $O^S$ but   
 for external, stand-by $O'$  the    $S',O^S$  evolution should be
 described by SLE as follows from  condition i).\\
     Below it will be argued that
this doublet state formalism (DSF)  corresponds to the measurements
description in Algebraic QM framework.

For the novel MS state:   
 $\Phi=|\phi^D, \phi^I\gg$ 
the dynamical component  $\phi^D$ is also
equal to QM density matrix  $\phi^D=\rho$ and obeys  to   SLE  :
\begin {equation}
    \frac {\partial \phi^D}{\partial  t} =[\phi^D,\hat {H}]  \label {AA8}
\end {equation}
  and the initial $\phi^D$ of (\ref{AAB}) evolves   
 at $t>t_1$ to $\phi^D(t)=\rho^p_{MS}$ of  (\ref{AA2}).
    $O$ information component  $\phi^I$ differs principally
from Breuer theory because it behaves stochastically
in the individual events. 
  Namely,  for $t \le t_0$ the initial
 $\phi^I=|O_0\rangle \langle O_0|$  - $O$ has
 no information on S at $t_0$. For
    the  final  $\phi^I(t)$ at $t \ge t_1$
   after the measurement
 at $t>t_1$  $\phi^I$ is the stochastic
state $\phi^I(n)=\phi^I_i$, where
$\phi^I_i=|O_i\rangle \langle O_i|$,
 with $i(n)$ described by  the probabilistic distribution with $P_i=|a_i|^2$.
  Therefore, such doublet, individual state $\Phi(n)$ 
can change from event to event  and $\phi^I$
is partly independent of $\phi^D$ being correlated with it only
 statistically.
 DSF $O$ subjective states $\phi^I$  can't  differ
the pure and mixed states with the same $|a_i|^2$. Therefore, Breuer
theorem conditions fulfilled and  the subjective state
collapse observed by $O$. 
  MS ensemble evolution described
 via the doublet statistical states  
$  |\Theta\gg=|\eta_D,\eta_I\gg  $,
where $\eta_D=\phi^D$, $\eta_I(t)$ describes 
the probabilistic distribution $\{P_i(t)\}$  of  $O$  $\phi^I_i$ observations
at given $t$.
%
Thereon, $\eta_I(t)$ defined by $\eta_D(t)$ which  obeys to SLE.
 Due to it,
 $\Theta$ evolution is reversible and the acquired $O$ information
can be erased completely. 
Naturally, 
the quantum  states for external $O'$ (and other observers)  also has
the same doublet form $\Phi'$. In the regarded situation
 $O'$ doesn't interact with MS  and so  $O'$ information
doesn't change after S measurement by $O$, eventually,
 for $O'$ MS evolution described by SLE only.
%


  Witnessing Interpretation proposed by Kochen (Kochen,1985) 
  is quite close to DSF but doesn't exploits the selfdescription
effects. It
 phenomenologically supposed that for apparatus $A$ ($O$ in our
notations) some S
measured value $Q$ in pure state always has random definite value
$q_j$ relative to $A$, yet 
  no new mathematical formalism different
from standard QM  wasn't constructed for its proof (Lahti,1990).


Plainly, in DSF $|O_i\rangle $ constitutes the  
 preferred basis (PB) in $\cal H_O$ and its appearance should be explained
in the consistent theory,
this problem   is well-known
 in standard QM with the Reduction Postulate (Busch,1995). 
In DSF  PB problem acquires the additional aspects related
to the information recognition by $O$. 
The plausible explanation prompts $O$ decoherence - i.e.
$O$ interaction with environment E (Zurek,1982; Guilini,1996). 
In this case the produced,  entangled S,$O$,E state  admits
 the unique,  orthogonal decomposition  which extracts $O$
 PB for the final MS states (Elby,1994).
 Tuning the interaction parameters,  PB can be made equivalent to 
$|O_i\rangle$ basis which resolves formally PB problem
(Mayburov,2002).  Despite its importance,
 for the simplicity
   the decoherence consideration and its influence on  
$O$ selfdescription omitted here. 
We plan to present this results in the forcoming paper,
 the preliminary calculations shows that the decoherence account
doesn't changes our principal conclusions
(Mayburov,2002).

\section {   Selfmeasurement in Algebraic QM}

Now   the quantum  measurements and $O$ selfdescription
will be regarded
in  Algebraic QM framework (Bratelli,1981).
  Besides the standard quantum effects,
Algebraic QM    describes successfully
the phase transitions and other nonperturbative phenomena 
which  standard QM fails to incorporate (Emch,1972). 
Consequently, there are the serious premises to regard Algebraic QM
as the consistent generalization of standard QM.
Algebraic QM was applied extensively
 to the superselection model of quantum measurements
when the detector D or environment E  regarded as the infinite systems
(Pimas,1990; Guilini,1996).
The algebraic formalism of nonperturbative QFT  
 was applied also to  the study of measurement 
dynamics in some realistic systems (Mayburov,1998).
%
%
In standard QM the fundamental structure is the fixed states set 
- Hilbert space $\cal H$ on which an observables - Hermitian operators
defined. Yet for some systems 
 the states set structure  principally differs
 from the arbitrary $\cal{H}$ and the standard QM
axiomatics becomes preposterous. 
In distinction, in Algebraic QM   the fundamental structure 
 is the  Segal algebra $\cal{U}$ of observables $A,B,...$
which incorporate the main properties of the studied system $S_f$
and eventually, defines
$S_f$ state set $\Omega$  (Emch,1972).
 Technically, it's more  convenient to  consider 
 $C^*$-algebra $\cal{C}$  for which $\cal{U}$ is the subset
and calculate $\cal{U}$ properties afterward.
 For our problems $\cal{C}$, $\cal{U}$ are in the unambiguous correspondence 
$\cal{C} \leftrightarrow \cal{U}$
and below their use is equivalent in this sense. 
%
   $S_f$   states set $\Omega$ defined by
$S_f$ $\cal{U}$    via the notorious
 GNS  construction; it demonstrates that
  $\Omega$ is the  vector space dual to
 the corresponding $S_f$ $\cal C$ (Bratelli,1981).
 Such states
called here the algebraic states $\varphi \in \Omega$ and
are defined as the normalized, positive, 
linear functionals on $\cal {U} $:
 $\forall A \in {\cal U}$;$ \; \forall \varphi \in \Omega$ it gives
$\bar{A}=\langle \varphi;A\rangle$.

Here only unitarily equivalent $S_f$ will be regarded;
 for  them $S_f$ $\varphi$ formally
corresponds   to QM density matrixes $\rho$ (Segal,1947).
%
The algebraic pure states are    $\Omega$  extremal points and they
  regarded as the algebraic individual states
 (AIS) $\xi$; their set denoted $\Omega^p$  (Emch,1972; Primas,1990).
%
The arbitrary $\varphi$ doesn't admit the  unambiguous
decomposition into AIS $\xi_i$ ensemble,
except the  situation when  $\varphi$ is pure; in this
case   $\xi=\varphi$.
 The algebraic mixed states $\varphi_{mix}$
can be constructed as $\xi_i$ ensembles;
  the   ensemble states $W_A$
  defined analogously to  the  described  QM ansatz.


  In many practical situations 
 only some restricted 
   linear subspace $\cal M_R$  or  subalgebra
 $\cal{U}_R$ of $S_f$ observables  algebra $\cal U$
is available for the observation.
 For such subsystems the restricted  algebraic states $\varphi_R$
 can be defined consistently
 via $A_R \in \cal{U}_R$ expectation values :
\begin {equation}
    \bar{A}_R=\langle \varphi;A_R\rangle=\langle \varphi_R;A_R\rangle
  \label {CC12}
\end {equation}
defining $\varphi \rightarrow \varphi_R$ restrictions;
their set denoted $\Omega_R$.
 $\varphi_R$ doesn't depend on any $A' \notin \cal{U}_R$,
therefore, $\forall \varphi_R,\,\langle \varphi_R;A'\rangle=0$ (Emch,1972).
For our MS only $O$ observables supposedly are available
for the observation (perception) and that makes the subalgebras
studies important for us.
Remind that  any classical system $S^c$  can be
described by some associative Segal algebra $\cal {U}^C$
 of $S^c$ observables $\{A\}$ (Emch,1972);
 in algebraic QM $\cal U$ associativity
corresponds to QM observables commutativity.
The theorem by Segal   proves that 
any  associative Segal  (sub)algebra $\cal{U}'$
is isomorphic to some algebra $\cal {U}^C$ of classical  
observables (Segal,1947); thereon, its  $\varphi^a$ states set
  $\Omega^a$ is isomorphic to
  the set $\Omega ^c$ of the  classical statistical states $\varphi^c$.
The corresponding AIS   - i.e. the pure states   
corresponds to the classical,
individual  states $\xi^c_i$ - points in $S^c$ parameters space.
For us the most important is the  case when
  $\cal{U}'$ includes only $I$ and single $A \ne I$; 
   there $\xi^c_i=\delta(q^A-q^A_i)$,  corresponding to 
 $A$ eigenvalues $q^A_i$ spectra.
 Consequently, even if quantum $S_f$ described by nonassociative $\cal{U}$,
 it contains the subalgebra $\cal{U}'\in U$
 ( and may be not unique)  for which
 the restricted AIS $\xi^c_i$ are classical with  the objective properties
$q^A_i$.
%



For the classical observing system $S^c_T$ described by some $\cal{U}^C$
its  selfmeasurement  $O$ restrictions are easy to find -
$O$ state depends only on $S^c_T$ coordinates $\{x^O_j\}$
 which are $O$ internal
coordinates (Breuer,1996). 
They constitute $\cal{U}^C_R$ subagebra of $\cal{U}^C$ but the realistic
 $O$ effective subalgebra $\cal{U}^C_O\in \cal {U}^C_R $  
 can be even smaller because
some $x^O_j$ can be uninvolved directly into the measurement process.  
QM  Correspondence principle  prompts that for the transition to
 the quantum case  $S^c_T \rightarrow S_T$
 $O$ restricted subalgebra $\cal{U}_R$ also
includes only $O$ internal observables. In quantum case
any effective subalgebra
 $\cal{U}_O \in \cal{U}_R$ stipulating
   the restricted states  sets $\Omega_R,\Omega_O$, correspondingly.   
 Our main hypothesis is
 that in any individual event to the arbitrary  $S_T$ AIS $\xi$ 
  responds  some  restricted $O$ AIS $\xi^O_j$. It 
 advocated below for $\cal U_O$,
for $\cal U_R$ it accepted $ad\;hoc$.



  MS  described by
 $ \cal{U}$ Segal algebra  for MS  observables  which defines
  $\varphi^{MS} \in\Omega$ properties.
 $O$  subalgebra is $\cal{U}_R$ which includes all $O$ 
internal observables. Then, $\varphi^R \in\Omega_R$
 is equivalent to $O$ QM statistical states $\rho$ set.
 Consequently, 
$O$ AIS set $\Omega^p_R$ is equivalent to $\cal{H}_O$  and
 any $O$  AIS $\xi^R_i $ corresponds to some  $O$
 state vector $ |O^r_i \rangle \in \cal H_O$.
We don't study here $\Omega_R$ states further,
note only that Breuer $O$ state (\ref {AA4}  )  $R_O \notin  \Omega^p _R$
and can't be AIS  on $\cal {U}_R$ for $a_{1,2} \ne 0$. 
%
To define   $\cal U_O$,  let's consider
  $\varphi^{MS} \rightarrow \varphi^O$ restriction properties.
 Remind that
  for the regarded MS dynamics of  (\ref {AA2}) $O$ 
 can measure only the  observable $Q_O$, 
for any other $Q'_O \neq F(Q_O)$  the final $\bar{Q}'_O=0$;
 it means
 $ \forall \varphi^O; \, \langle  Q'_O ; \varphi^O\rangle=0$
for $O$  restricted, algebraic states.
%
%
%
From that follows that 
$\cal{U}_O \in \cal {U}_R$  effective  $O$  subalgebra
 includes only  $Q_O$ and $I$.
Really, only in this case 
 $ \forall \varphi' \in \Omega_O; \, \langle  Q'_O ; \varphi'\rangle=0$;
 each $\varphi^O$ corresponds to $\varphi'$ with the same $\bar{Q}_O$
and vice versa.
 Therefore,  $\varphi^O$ set $\Delta_O$ is isomorphic to $\Omega_O$.
 There is no other $\cal U_R$ subalgebras with such properties and
  that settles $\cal U_O $ finally. Therefore,
obtained $\varphi^O$ are  equivalent to $R_O$,  in agreement with 
  MS the statistical states $\rho_{MS}$ restriction
to $O$ which are equal to $R_O$ of (\ref{AA4}) as was shown above.
%
From Segal theorem  for  $\cal{}U_O$  
 the restricted algebraic $O$ 
 states $\varphi^O\in \Omega_O$   are isomorphic to classical,
probabilistic $q^O_i$ distributions,
 $O$ AIS $\xi^O$
 are  isomorphic to the classical, pointlike  states:
$$
            \xi^O_i= \delta(q^O-q^O_i)
$$
for $Q_O$ eigenvalues.  
%
  For  the incoming S state
$\psi_s=|s_i\rangle$  results in 
 $\Psi^{MS}_i=|s_i\rangle|O_i\rangle$ which are $\Omega$ extremal points,
 $O$ restricted states $\varphi^O_i =  |O_i\rangle\langle O_i|$
  are $\Omega_O$ extremal point  and   AIS $\xi^O_i=\varphi^O_i$.
In any $Q$  eigenstate $|s_i\rangle$   measurement 
 the final MS restricted
 state  from  $O$ 'point of view' describes the definite 
$Q_O$ value   $q^O_i$ which establishes
operationally $\xi^O_{i,j}$ distinction in the individual events.

  For the incoming S mixture with the $|s_i\rangle$ probabilities $|a_i|^2$
 MS algebraic final state is
  $\varphi_{mix}=\rho^m_{MS}$ of  (\ref {AA3}); 
%
%
 the corresponding $O$ restricted state  $\varphi^O_{mix}$
 defined from the relation for $\bar{Q}_O$ :
$$
    \bar{Q}_O=\langle\varphi^O_{mix} ;Q_O\rangle=
    \langle\varphi_{mix};Q_O\rangle=\sum |a_i|^2 q^O_i
$$
which results in the solution  $\varphi^O_{mix}=\sum  |a_i|^2\varphi^O_i$.
From the regarded correspondence
of MS $\xi^{MS}$ and $O$ AIS $\varphi^O_{mix}$
 represents the stochastic mixture of AIS $\xi^O_i$ 
 described by $O$  ensemble state 
  $W^O_{mix}=\{ \xi^O_i;\,P_i=|a_i|^2;\,i=1,2\}$. If  the incoming  S state  
 is $Q$ eigenstates superposition
$\Psi_{MS}^{in}$ of (\ref{AAB}),  MS final algebraic state
$\varphi^{MS}$ with the same $|a_i|^2$ results in  the same $\bar{Q}_O$ value.
Therefore, its  restricted algebraic state
coincides with the mixed one $\varphi^O=\varphi^O_{mix}$.
From   Segal theorem 
 in $\Omega^p_O$ all $O$  individual states are AIS $\xi^O_i$,  
possesing the definite properties $q^O_i$.
 There is no
other individual states $\xi^O_a \ne \xi^O_{i}$, consequently,
 MS restricted AIS  in each event can be only one of $\xi^O_i$.
 Eventually,
 if MS state is $\Psi^{MS}_j$ superposition of (\ref {AA2}),
 to dispatch the correct $\bar{Q}_O$ for $\xi^O$ ensemble,
$\xi^O_i$ should appear at random  with the probabilistic distribution $P'_i$
defined by $\bar{Q}^O$.
Yet for such $W^O$ content the only solution which results in the
necessary $\bar{Q}_O$ value is  $P'_i=|a_i|^2$ and so
 $W^O=W^O_{mix}$.
 It demonstrates
  that $\xi^{MS}\rightarrow \xi^O$ restriction map is stochastic. 

In general, any two physically different states operationally
discriminated by the particular
 observation procedure which reveals this states
difference via the  difference of some observables values
distributions.
For the statistical states it demonstrated by their probabilistic
distributions parameters, for the individual states $\xi_{a,b}$ 
such difference can be extracted from some observables $A,B$
 eigenvalues $q^{A,B}_{i,j}$ for which this states are the eigenstates.
 In that case   this values 
 can be obtained and compared in the single event per each state
(Mittelstaedt,1998). 
In our case the only $\cal{U}_O$ observable is $Q_O=\sum q^O_i P^O_i$ and
$\xi^O_{i,j}$ difference reflected by $q^O_{i,j}$ difference. 
 If to assume that some other $\xi^O_a \ne \xi^O_i$ exists,
it needs also some other observable $Q^e \ne F(Q^O);\, Q^e$ should
belong to  $ \cal{U}_O$ to differ it from $\xi^O_i$, 
but it's inconsistent with the obtained $\cal{U}_O$ structure.
 In particular, Breuer restricted state $R_O$ of (\ref {AA4}) analog
$\xi^O_R=\sum |a_i|^2 \xi^O_i$  for $a_i \ne 0$
can't be $O$ individual state on $\cal{U}_O$ because it
  isn't $\Omega_O$ extremal point.
 Moreover, the arbitrary $\varphi^O$ admits the unique
decomposition into $\xi^O_i$ set and  can be interpreted as $\xi^O_i$ 
ensemble  with the given probabilities.
%
  Since $\xi^O_i \in \Omega _R$, 
it can be taken  also as the  possible ansatz for MS  states  restriction on 
   $\cal {U}_R$.
DSF doublet state $\Phi$ components $\phi^D,\phi^I$
 are equivalent to $\xi^{MS}$, $\xi^O$ correspondingly.

If to analyze this results  from the Information-Theoretical premises,
 note that
 the difference between the pure and
mixed MS states reflected by $B$ IT of (\ref {AA5}) expectation values.
Therefore, $O$ possible observation of S pure/mixed $W^O$
 states difference means
that $O$ can acquire the information on   $B$  expectation value.
 But $B \notin \cal {U}_R$ and isn't correlated with $Q_O$ via
S,$O$ interaction alike $Q$ of S; so this assumption  is prepostereous.
%
%
 Note that MS individual states $\xi^{MS}$ symmetry
 is larger than the symmetry   of the restricted $O$ states.
In Algebraic QM such symmetry reduction results in
 the phenomena of Spontaneous Symmetry Breaking, by the
analogy the  discussed  randomness apearence can be
called Information Symmetry Breaking.
 In practice
it's possible that $O$ effective subalgebra is larger then $\cal U_O$
 but this case
will demand more complicated calculations which we plan to present
in the forcoming papers.
 In Algebraic QM 
the only important condition for the classicality appearance
is $\cal{U}_O$ observables commutativity and it's reasonable to expect
 it to be feasible also for complex IGUS structures.


Despite  of the acknowledged  Algebraic QM achievements,
 its foundations are still
discussed and aren't finally established. In particular, 
it's still unclear whether all the algebraic states corresponds to the
physical states (Primas,1983). This questions are  important by themself 
and  are essential for our formalism feasibility. By our choice
of the initial MS states we avoid it in the regarded model.
In particular, we admitted  without proof
 that for MS AIS $\xi^{MS}$  some   $O$ restricted AIS responds
in any event. It agrees with the restricted states consideration 
 as the real physical states, 
 but on the whole, this assumption needs further clarification.



For the conclusion, the
 information-theoretical restrictions on the quantum measurements
were  studied  on the simple selfdescription  model of IGUS $O$.
Breuer selmeasurement study shows that
by itself  the $O$ inclusion  as the quantum object
into the measurement scheme doesn't result in the
 state collapse appearance (Breuer,1996).
Our considerations indicates that to describe the
 state collapse and in the same time to conserve
Schr$\it\ddot{o}$dinger linear evolution, it's necessary 
 to extend the quantum states set over standard QM Hilbert space.
Such modification proposed in DSF involving the doublet states $\Phi$,
where one of its componenets $\phi^I$ corresponds to $O$
 subjective information - i.e. $O$ selfdescription.
  Algebraic QM  
presents the additional arguments in favour of this approach,
 in its   formalism $O$ structure
  described by $O$ observables algebra $\cal {U}_O$
which defines the multiplet states set analogous to $\Phi$. 
In Algebraic formalism the stochastic events appearance
 stipulated by  MS individual states restriction to $O$.
In our opinion  the obtained results evidence
 that it's impossible to solve  the Measurement Problem
 without accounting of the information system 
$O$ interactions at quantum level and its information
acquisition restrictions (Zurek,1998).
\\
\\
\qquad \qquad \qquad {\Large { References}}\\
\\
 (1981) Y.Aharonov, D.Z. Albert Phys. Rev. D24, 359 
\\
 (2000) G.Bene, quant-ph 0008128
\\
 (1979) O.Bratteli, D.Robinson 'Operators Algebra and
Quantum Statistical Mechanics' (Springer-Verlag, Berlin)
\\
 (1996) T.Breuer, Phyl. of Science 62, 197 (1995),
 Synthese 107, 1 (1996)
\\
 (1996) P.Busch, P.Lahti, P.Mittelstaedt,
'Quantum Theory of Measurements' (Springer-Verlag, Berlin,1996)
\\
 (1990) W. D'Espagnat, Found Phys. 20,1157,(1990)
\\
%
 (2002) R.Duvenhage, Found. Phys. 32, 1799  
\\
 (1994) A.Elby, J.Bub Phys. Rev. A49, 4213 
\\
 (1972) G.Emch, 'Algebraic Methods in Statistical Physics and
Quantim Mechanics',\\
 (Wiley,N-Y) 
\\
 (1988) D.Finkelstein, 'The Univrsal Turing Machine:
 A Half Century Survey', (ed. R.Herken, University Press, Oxford) 
\\
 (1996) D.Guilini et al., 'Decoherence and Appearance of
Classical World', (Springer-Verlag,Berlin) 
\\
 (1985) S.Kochen 'Symposium on Foundations of Modern Physics'
  , (World scientific, Singapour)
\\
 (1990) P. Lahti Int. J. Theor. Phys. 29, 339 
\\
 (1939)  London F., Bauer E. La theorie de l'Observation
 (Hermann, Paris)   
\\
 (1998) S.Mayburov, Int. Journ. Theor. Phys. 37, 401 
\\
 (2001) S.Mayburov  Proc. V  QMCC Conference, Capri, 2000,
(Kluwer, N-Y); $ \setminus$quant-ph 0103161
\\
 (2002) S.Mayburov Proc. of Vth Quantum Structures conference,
Cesenatico, 2002; Quant-ph 0205024; Quant-ph 0212099
\\
 (1998) P.Mittelstaedt 'Interpretation of
Quantum Mechanics and Quantum Measurement\\ Problem',
(Oxford Press, Oxford)
\\
 (1932) J. von Neuman 'Matematische Grunlanden
 der Quantenmechaniks' , (Berlin)
\\
 (1983) H.Primas,  'Quantum Mechanics,
 Chemistry and Reductionism' (Springer, Berlin)
 (1990) H.Primas,  in  'Sixty two years of uncertainty'
,ed. E.Muller, (Plenum, N-Y)
\\
 (1995) C. Rovelli, Int. Journ. Theor. Phys. 35, 1637; 
quant-ph 9609002  
\\
 (1947) I.Segal, Ann. Math., 48, 930    
\\
 (1993) K.Svozil 'Randomness and undecidability in Physics',
(World Scientific, Singapour)
\\
%
 (1961) E.Wigner,  'Scientist speculates',(Heinemann, London)
\\
 (1982) W.Zurek, Phys Rev, D26,1862 
\\
 (1998) W.Zurek Phys. Scripta , T76 , 186 
\\
\\
\end {document}